\documentstyle[11pt,aaspp4]{article}

\newcommand{\gsim}{{}^>_\sim}

\begin{document}

\title{The Influence of the Photoionizing Radiation Spectrum on Metal~-~Line
Ratios in Ly$\alpha$ Forest Clouds} 

\author{ Mark L. Giroux and J. Michael Shull\altaffilmark{1}}

\affil{Center for Astrophysics and Space Astronomy, \\
Department of Astrophysical, Planetary, and Atmospheric Sciences \\
University of Colorado, Campus Box 389, Boulder, CO 80309 }

\altaffiltext{1}{also at JILA, University of Colorado and National Institute of 
Standards and Technology}

\begin{abstract}

Recent measurements of Si~IV/C~IV ratios in the high-redshift Ly$\alpha$ forest
(\markcite{SC96}Songaila \& Cowie, AJ, 112, 335 [1996a]; \markcite{S97}Savaglio
et al., A\&A, in press [1997]) have opened a new window on chemical enrichment
and the first generations of stars. However, the derivation of accurate Si/C
abundances requires reliable ionization corrections, which are strongly
dependent on the spectral shape of the metagalactic ionizing background and on
the ``local effects'' of hot stars in nearby galaxies.  Recent models have
assumed power-law quasar ionizing backgrounds plus a decrement at 4 Ryd to
account for He~II attenuation in intervening clouds.  However, we show that
realistic ionizing backgrounds based on cosmological radiative transfer models
produce more complex ionizing spectra between 1--5 Ryd that are critical to
interpreting ions of Si and C. We also make a preliminary investigation of the
effects of He~II ionization front non-overlap. Because the attenuation and
re-emission by intervening clouds enhance Si~IV relative to C~IV, the observed
high Si~IV/C~IV ratios do {\it not} require an unrealistic Si overproduction
[Si/C $\geq 3$ (Si/C)$_{\odot}$]. 
If the ionizing spectrum is dominated by
``local effects'' from massive stars, even larger Si~IV/C~IV ratios are
possible. However, unless stellar radiation dominates quasars by more than a
factor of 10, 
we confirm the evidence for some Si overproduction by massive
stars; values
Si/C $\approx 2$ (Si/C)$_{\odot}$
fit the measurements
better than solar abundances. Ultimately, an adequate interpretation of the
ratios of C~IV, Si~IV, and C~II may require hot, collisionally ionized gas in a
multiphase medium. 

\end{abstract} 


\section{Introduction}

The detection of metals associated with the ``high'' column density [N(H~I) $>
10^{14.5}$ cm$^{-2}$] Ly$\alpha$ forest of absorbers in quasar spectra
(\markcite{C95}Cowie et al. 1995; \markcite{T95}Tytler et al. 1995) provides
new challenges to the understanding of the nature and origin of these
absorbers. The contamination by metals must now be accounted for in models for
cloud evolution. It is also possible to perform a detailed analysis of the
thermodynamic and ionization state of these absorbers, which have previously
been probed only by their hydrogen Ly$\alpha$ absorption and the integrated
effect of their He~II Ly$\alpha$ absorption 
(\markcite{GFS95}Giroux, Fardal, \& Shull 1995). If
the clouds are photoionized, the detection of C~II, C~IV and Si~IV may
constrain the shape of the metagalactic radiation spectrum and indicate the
epoch of dramatic changes in the ionizing radiation field at $z > 3$
(\markcite{SC96}Songaila \& Cowie 1996a; \markcite{S97}Savaglio et al. 1997). 

The multiple C~IV lines associated with a single H~I absorption complex
\markcite{SC96}(Songaila \& Cowie 1996a) belies a simple picture of a
homogeneous cloud. These absorbers, like Lyman Limit systems, may better be
interpreted with multiphase models (\markcite{GSS94}Giroux, Sutherland, \& Shull
1994; \markcite{PRR96}Petitjean, Riediger, \& Rauch 1996;
\markcite{HRS97}Haehnelt, Rauch, \& Steinmetz 1997;
\markcite{H97}Hellsten et al. 1997) that may include hot,
collisionally ionized gas. We defer such interpretation to a later study,
focussing here on the tradition of analyzing metal-line ratios with
single-phase photoionization models.  In this way, we can isolate the effects of
the incident spectrum on species with ionization potentials in the range 1--5
Rydbergs (e.g., C~III, C~IV, Si~III, Si~IV). For example,
\markcite{SC96}Songaila \& Cowie (1996a) argue that the observed increase in
the Si~IV/C~IV ratio at $z > 3.1$ suggests that the He~III regions around
ionizing sources may not have overlapped by that epoch, cutting off all
radiation with $h \nu > 54.4$ eV. Here, we primarily consider models for the
averaged metagalactic background in which He~II ionization fronts have
overlapped before $z \approx 4$. We do consider a limiting case in which no
photons above 4 Ryd are present in the background, as well as a case where we
restore the much less heavily attenuated x-ray background. 

In the past decade, metal-line 
ratios in Lyman Limit ($N_{HI} {}^>_\sim 10^{17}$ cm$^{-2}$)
absorbers have been used to probe the shape of the ionizing spectrum,
and to estimate the intrinsic metallicity of the absorbers 
(cf.  \markcite{BS86}Bergeron \& Stasinska 1986; 
\markcite{SBS88}Sargent, Boksenberg \& Steidel 1988;
\markcite{SSB88}Steidel, Sargent \&, Boksenberg 1988;
\markcite{SS89}Steidel \& Sargent 1989;
\markcite{BI90}Bergeron \& Ikeuchi 1990;
\markcite{MO90}Miralda-Escud\'e \& Ostriker 1990; 
\markcite{DS91}Donahue \& Shull 1991;
\markcite{M92}Madau 1992).
With the advent of new spectrographs on large telescopes, the lower $N_{HI}$
absorbers can be exploited in the same way. These absorbers hold a special
interest, because the source of their metal enrichment is still unclear. While
the metal lines in Lyman limit absorbers have been associated with the haloes
of bright galaxies, the number of Ly$\alpha$ forest absorbers greatly exceeds
the corresponding number of bright galaxies, and the Ly$\alpha$ absorbers do
not seem to show the same clustering properties as galaxies. 

As yet, metal lines have only been associated with ``high-column'' absorbers
[N(H~I) $>10^{14.5}$ cm$^{-2}$], which numerically are a small fraction of the
Ly$\alpha$ forest. It is still possible that these clouds may represent an
overlap of the high-column end of the pristine Ly$\alpha$ forest clouds and the
low-column end of the metal enriched clouds associated with galactic haloes.
As a result some, if not all, of the metal-line absorbers may
still be associated with the galaxies that contain the stars responsible for
their metals.  For example, over a Hubble time at $z = 3$, galactic outflows at
$300 V_{300}$ km~s$^{-1}$ will transport heavy elements a distance
\begin{equation}
   d \leq (300~{\rm kpc}) V_{300} h^{-1} \left[ \frac {1+z}{4} \right]^{-3/2}
\end{equation}
for a Friedmann universe with $H_0 = (100~{\rm km~s}^{-1})h$ and $\Omega_0 = 1$.
In practice, the period of heavy element injection may have lasted only $10^8$
yrs, and the metals could move distances of only 30-50 kpc from their sources
(bright galaxies, dwarf galaxies, globular clusters). 

As \markcite{MS96}Madau \& Shull (1996) have emphasized, metal enrichment of
the Ly$\alpha$ absorbers implies a substantial Lyman continuum (LyC) emission
accompanying the star formation at $z {}^>_\sim 3.5$. Thus, a large fraction of
the metagalactic background at high redshift could be due to massive stars, a
point also made by \markcite{GS96}Giroux \& Shapiro (1996) and 
\markcite{S97}Savaglio et al.
(1997). Furthermore, the absorbers with metal lines may be sufficiently close
to the local sources that the incident radiation field is dominated by local
hot-star radiation. In contrast, if the metal enrichment of the Ly$\alpha$
forest arises from Population III stars at $z > 10$ (\markcite{C85}Couchman
1985; \markcite{OG96}Ostriker \& Gnedin 1996), the stellar ionizing radiation
will be greatly attenuated by intergalactic absorption, and the metagalactic
background at $z = 3$ will be dominated by the harder spectrum of quasars. 
Because metal-line systems would then be less likely to be associated with
nearby sources of stellar radiation, their absorption-line ratios would be more
representative of the metagalactic ionizing background. 

Preliminary analyses (\markcite{SC96}Songaila \& Cowie 1996a;
\markcite{S97}Savaglio et al. 1997; \markcite{H97}Haehnelt 1997) of the
Ly$\alpha$ forest clouds have assumed photoionization by simple power-law
ionizing spectra. The only way these models account for filtering of the
ionizing source spectrum by intervening clouds is to include a decrement at
$\nu_{HeII}$ (4 Ryd). More realistic models include several additional effects.
First, when absorption due to intervening clouds is included, the background
spectrum just above the H~I and He~II ionization edges shows a flatter power
law than that of the sources (\markcite{MO90}Miralda-Escud\'e \& Ostriker 1990;
\markcite{MM94}Madau \& Meiksin 1994; \markcite{GFS95}Giroux et al. 1995).
Second, \markcite{HM96}Haardt \& Madau (1996) and \markcite{FGS97}Fardal,
Giroux, \& Shull (1997) show, using cosmological radiative transfer, that cloud
emission by recombination to the H~I and He~II ground states and He~II
Ly$\alpha$ and 2-photon radiation alters the extent of the decrement at
$\nu_{HeII}$ and changes the shape of the spectrum between 1 and 5 Rydbergs.
Including a more realistic spectral shape will alter the relative populations
of metal ions with edges in the 1 to 5 Ryd range even if the ionizing sources
are all assumed to be quasars with intrinsic power-law ionizing spectra. 

In this paper, we consider the extent to which these alterations of the assumed
shape of the incident radiation field affect the interpretation of the
metallicity of these absorbers and the nature of the photoionizing sources. We
use the data sets of \markcite{SC96}Songaila \& Cowie (1996a) and
\markcite{S97}Savaglio et al. (1997), which include Si~IV and C~IV lines
associated with higher column density Ly$\alpha$ forest clouds.
\markcite{SC96}Songaila \& Cowie (1996a) provide an important subsample for
which C~II lines are also observed. 

\section{The Metagalactic Radiation Background}

\subsection{Ionizing Sources}

Assumptions about the nature of the ionizing sources of the metagalactic
background are uncertain, since the intrinsic EUV spectra of quasars are
not well defined. For simplicity, we assume that all quasars have constant
power-law spectra $F_\nu \propto \nu^{-\alpha_s}$. Our choice of spectral
index, $\alpha_s = 1.8$, is consistent with HST observations of the EUV
spectrum of quasars (\markcite{Z97}Zheng et al. 1997) for radio-quiet quasars
at $z = 1.5-2.0$. As discussed by \markcite{FGS97}Fardal et al. (1997), if the
ionizing background (1-5 Ryd) is dominated by these quasars, the He~II
Ly$\alpha$ absorption optical depth $\tau_{HeII} \approx 1$ at $z = 3$.  For
clarity, we denote the cloud-filtered, quasar-dominated mean intensity by
$J_{Q,f}(\nu)$. 

The characteristic spectrum of stellar sources in starburst galaxies is also
not well known. Most models agree (e.g., \markcite{BC93}Bruzual \& Charlot
1993) that few photons are present with frequency above 54.4 eV, and that the
spectrum falls off above $h\nu \approx$ 45 eV.  
Recent line-blanketed, non-LTE hot-star models (\markcite{G92}Gabler et al.
1992; 
\markcite{S96}Sellmaier et al. 1996;
\markcite{SdK}Schaerer \& de Koter 1997) 
vary considerably in the amount of
radiation in the He~I continuum ($24.6 < h\nu < 54.4$ eV) 
they predict, compared to the LTE
line-blanketed models of 
\markcite{K92}Kurucz (1992). 
An ongoing uncertainty, for our purposes, is the lack
of low-metallicity stellar models, which are more appropriate
for high-redshift starburst galaxies.
For the present work, we use a
starburst spectrum (\markcite{SS97}Sutherland \& Shull 1997) discussed in the
context of the ionizing spectrum of metal-forming galaxies by
\markcite{MS96}Madau \& Shull (1996) [see their Fig. 3]. In brief, it is a time
integrated spectrum of a Gaussian starburst of width $2 \sigma = $ 4 Myr
containing $5000 M_\odot $ of stars between 8 and 85 $M_\odot$ with IMF slope
$\Gamma = 1.6$ where $dN(>M)/dM \propto M^{-\Gamma}$ and $N(>M)$ is the number
of stars with mass greater than $M$. 

Estimates of the mean metagalactic ionizing intensity at $z > 2$, based on the
statistical diminution in the number of Ly$\alpha$ forest lines close to
quasars (the proximity effect), indicate a specific intensity at the hydrogen
Lyman limit, $\nu_H$, of $J_{-21} = 0.7-1.5$ where $J_{-21}$ is in units of
$10^{-21}$ergs~cm$^{-2}$ s$^{-1}$ Hz$^{-1}$ sr$^{-1}$ (\markcite{CEC97}Cooke,
Espey, \& Carswell 1997). If all Ly$\alpha$ absorbers possess a metallicity of
$Z = 0.01 Z_\odot$ by redshift $z \approx 3.5$, the stellar ionizing photons
associated with the production of the metals may be comparable to the ionizing
photons produced by quasars (\markcite{MS96}Madau \& Shull 1996;
\markcite{GS96}Giroux \& Shapiro 1996). If these ionizing photons are produced
at higher redshift (e.g., Population III stars at $z > 5$) they will be highly
attenuated by IGM absorption at $z > 4$. A further uncertainty is the fraction,
$\langle f_{\rm esc} \rangle$, of ionizing photons that successfully escape
from the galaxy in which the stars reside (\markcite{DS94}Dove \& Shull 1994).
It does seem possible, though, that stellar sources could be an important
contributor to the metagalactic background, which could dominate in regions
local to starburst galaxies. 

For example, the total Lyman continuum production rate of the Milky Way was
recently inferred from COBE observations of [N~II] 205 $\mu$m emission to be
$3.5\times 10^{53}$ photons~s$^{-1}$ \markcite{B94}(Bennett et al. 1994), and
most starburst galaxies exceed the value $S = (10^{54})S_{54}$
photons~s$^{-1}$. Assuming that a substantial fraction, $\langle f_{\rm esc}
\rangle \geq 0.1$, of these photons escape these starburst galaxies, the mean
photoionization rate of hydrogen due to the galactic source may exceed $J_{-21}
= 1$ if the absorber lies within a distance 
\begin{equation}
  R_{\rm cr} = (88~{\rm kpc}) \; S_{54}^{1/2} J_{-21}^{-1/2} \;
    \langle f_{\rm esc} \rangle ^{1/2} 
    \left[ \frac {\alpha_s +3}{4.8} \right]^{1/2} \; .
\end{equation}
Here, we have assumed the QSO spectra in the range 1--5 Ryd to be power laws 
with spectral index $\alpha_s \approx 1.8$.  

In this paper, we account for a stellar contribution to the mean intensity
falling upon our absorbers in a global way.  We assume that up to $2/3$ of the
ionizing emissivity from primary sources is stellar, as opposed to secondary
sources due to recombination radiation from clouds. The remainder of the
primary emissivity is then from AGN. We then compute the radiative transfer of
primary and secondary sources through the IGM to derive the metagalactic
ionizing background, $J_{SQ,f}(\nu)$. The mean intensities $J_{Q,f}$ (pure QSO
sources) and $J_{SQ,f}$ (stars + QSOs) therefore bracket our estimates of the
contribution of stellar sources to the uniform metagalactic background,
$J_{MB}(\nu)$. However, even if the metagalactic ionizing background is quasar
dominated, the absorber may primarily see a {\it local} unfiltered source of
stellar photons, $J_{L} (\nu)$. We consider models which assume $J_{L} / J_{MB}
= 5, 10, 20$ at $\nu = \nu_H$ for $J_{MB} = J_{Q,f}$ and $J_{L} / J_{MB} = 3,
5, 10, 20$ for $J_{MB} = J_{SQ,f}$. 

\subsection{He~II Ionization Fronts}

Within a picture of ionizing sources turning on in a neutral
medium,
\markcite{MR94}Miralda-Escud\'e \& Rees (1994)
made the important suggestion that
the He~II I-front
which propagates away from sources 
with a sufficiently soft ionizing spectrum
can lag the hydrogen I-front.
\markcite{MM94}Madau \& Meiksin (1994) showed that,
if $\alpha_s = 1.9$, the overlap of He~II I-fronts could be
delayed to as late as $z=3$.  As a result, both
\markcite{J94}Jakobsen et al. (1994) and
\markcite{SC96}Songaila \& Cowie (1996) suggested
that, at $z = 3.1$, He~II I-fronts around the 
dominant ionizing sources may not have overlapped.
Although a full
treatment of overlapping He II I-fronts is beyond the scope of this paper, we
can make a simple estimate of the ratio of I-front velocities for H~I and
He~II.  From the flux-limiting equations governing I-front propagation
(\markcite{SG87}Shapiro \& Giroux 1987; \markcite{DS87}Donahue \& Shull 1987),
the velocity relative to the local Hubble flow is given by 
\begin{equation}
   \left[ \frac {dr_i}{dt} - H r_i \right] = 
        \frac {S_i} {4 \pi r_i^2 n_i}   \; ,
\end{equation}
where $i$ refers either to the H~I ionizing front (1 Ryd continuum) or to the
He~II ionizing front (4 Ryd continuum).  (The He~I ionizing front is assumed to
be coincident with the H~I front, and we neglect recombinations
and attenuation within the ionized zone.)  
Here, $S_i$ (photons s$^{-1}$) represents
the photon production rates of the source in the H~I and He~II continua, $n_i$
is the density of H~I or He~II, and $r_i$ is the front distance measured from
the source. 

The propagation speeds of the H~I and He~II ionization fronts will be governed
by the ``flux-to-density'' ratio, $(S_i/n_i r_i^2$), at 1 and 4 Ryd. For a QSO
spectrum with power-law index $\alpha_s$, one finds that $S_{\rm HeII}/S_{\rm
HI} = 4^{-\alpha_s}$. 
The primordial helium abundance by mass has been estimated at
values $Y_P=0.231 \pm 0.006$
\markcite{SK93}(Skillman \& Kennicutt 1993)
and 
$Y_P = 0.232 \pm 0.003$ 
\markcite{OS95}(Olive \& Steigman 1995), while
\markcite{C94}Copi et al. (1994)
have suggested
a $2\sigma$ concordance range of
$Y_P = 0.221-0.243$ from theoretical models of Big-Bang 
nucleosynthesis. 
If helium has a cosmological abundance, $n_{\rm He}
/ n_{\rm H} = 0.0785$ by number ($Y = 0.239$ by mass), the flux-to-density
ratios at a fixed radius $r_i$ are equal for 
\begin{equation}
 \alpha_s = - \frac {\ln [n_{\rm He}/n_{\rm H}]} {\ln 4} \approx 1.84   \; .
\end{equation} 
At this critical spectral index, the H~I and He~II ionization fronts will
coincide. Interestingly, this critical index is very close to the mean index
found by 
\markcite{Z97}Zheng et al. (1997) 
for radio-quiet quasars.  For AGN with harder
spectra ($\alpha_s < 1.84$), the He~II ionization front will precede the
H~I front by a small amount, equivalent to a few optical depths in the
4 Ryd continuum.  However, for those AGN with softer spectra
($\alpha_s > 1.84$), the He~II front will propagate at a lower speed and
lie well within the H~I front.  For example, 
\markcite{Z97}Zheng et al. (1997) found that
radio-loud quasars had a mean index $\alpha_s \approx 2.2$.  In this case,
the regions outside the He~II fronts will see no 4 Ryd continuum radiation.

We therefore consider a limiting case in which all photons above 4 Ryd are
removed from the metagalactic background spectrum.  We also consider an
intermediate case, arising from the fact that quasars emit radiation at least
into the x-ray range of energies, so that high-energy photons will not be
strongly attenuated, even in regions where helium is entirely in the He~II
ionization stage.  In the Ly$\alpha$ clouds, we estimate the effect of the
presence of these high-energy photons by presuming that H II regions have
overlapped by $z=4$, and that the Ly$\alpha$ forest clouds, whose distribution
in $N_{\rm HI}$ is known (see \S 2.2), have a maximal He~II/H~I ratio. In
interpreting He~II Gunn-Peterson measurements, several groups 
(\markcite{MM94}Madau \& Meiksin
1994; \markcite{GFS95}Giroux et al. 1995) 
find that the ratio, $\eta$ = N(He~II)/N(H~I), must
be 50--100 in order to explain the $\lambda < 304$~\AA\ absorption as He~II
line opacity in the H~I Ly$\alpha$ forest.  This large ratio arises because
He~II is more difficult to photoionize than H~I. 

The H neutral fraction $f_{HI}$ is always less than $(n_{\rm He}/n_{\rm H})
\eta^{-1}$, where $\eta =$ N(He~II)/N(H~I) within an absorbing cloud. Thus, the
maximal ratio $\eta_{max}$ is bounded by limits on the baryon density from Big
Bang nucleosynthesis (BBN), $\Omega_b h^2 \le 0.024$ (\markcite{C94}Copi et al.
1994). At $z=3.4$, if only absorbers with $N_{HI} < 10^{15}$ cm$^{-2}$ are
considered to contribute to $\Omega_{HI}$, we find $\eta_{max} < 5000$. This
limit produces a local continuum optical depth at 4 Ryd, $d\tau / dz > 180$ at
$z =3.4$, falling to $4$ by $\nu = 25\nu_H$, the same continuum optical depth
as that at 1 Ryd. As a result, for this model with non-overlapping He~II
I-fronts, we only remove photons between 4 and 25 Ryd, retaining metagalactic
radiation background for $\nu > 25 \nu_H$.  This is a more conservative
estimate than if all He is assumed to be in He~II and distributed uniformly. 

\subsection{Cloud Opacity/Emission}

For a random spatial distribution, the local continuum optical depth of the
intervening absorbers  is given by (\markcite{PMB80}Paresce, McKee, \& Bowyer
1980) 
\begin{equation}
   { {d\tau(\nu)} \over {dz} } = 
   \int^\infty_0 {\partial^2N_c \over {\partial N_{H I} \partial z} }
   {[1 - e^{-N_{HI}\sigma_{eff}(\nu)}]} \; dN_{H I} \; , 
\end{equation}
where, for clouds composed of H and He,
\begin{equation}
   \sigma_{eff}(\nu) = \sigma_{HI}(\nu) +
   \left[ {N_{HeI} \over N_{HI}} \right]  \sigma_{HeI}(\nu) +
   \left[ {N_{HeII} \over N_{HI}} \right] \sigma_{HeII}(\nu) \; ,
\end{equation}
and ${{\partial^2 N_c} \over {\partial N_{HI} \partial z}}$ is the distribution
of absorbers in column density and redshift. This distribution has often been
parameterized by a function of the form 
\begin{equation}
{{\partial^2 N_c} \over {\partial N_{HI} \partial z}} =  A
{(1+z)}^\gamma N_{HI}^{-\beta} \;
\end{equation}
for $N_l$(H~I) $\leq N_{HI} ({\rm cm}^{-2}) \leq N_u$(H~I). Using new line
lists from {\it Keck} data, \markcite{FGS97}Fardal et al. (1997) have completed
a review of the statistics available on Ly$\alpha$ absorbers, Lyman Limit
Systems, and damped Ly$\alpha$ systems. They suggest that the distribution of
absorbers is best represented by a broken power law in $N_{HI}$.  If 
\begin{equation}
{{\partial^2 N_c} \over {\partial N_{17} \partial z}} =  A_i
{(1+z)}^\gamma N_{17}^{-\beta_i} \; , 
\end{equation}
where $N_{17}= N_{HI} / (10^{17}$~cm$^{-2})$, $\gamma = 2.585$, and $\beta_i$
and $A_i$ represent coefficients appropriate for a given range in $N_{17}$ (see
Table 2). We adopt their Model 2 for this paper.  Use of either of their other
two models does not affect our conclusions. 

Just as absorption due to intervening clouds will strongly alter the background
radiation spectrum, the re-emission of ionizing radiation from the clouds will
also affect the radiation (\markcite{HM96}Haardt \& Madau 1996; 
\markcite{FGS97}Fardal et al.
1997). Recombinations in the clouds produce continuum radiation (H~I Lyc, He~II
Lyc, He~II Bac, and He~II 2-photon) as well as He~II Ly$\alpha$ ($\lambda$304)
line radiation. Accounting properly for the Lyc and Bac radiation is critical to
the determination of the level of the mean intensity above 4 Ryd. While the
line and $2$-photon radiation are not important for that, they strongly alter
the spectrum between 1 and 3 Ryd and affect the populations of ions with
thresholds in that range (see Table 1). A complete discussion of the treatment
of reemitted radiation used in computing the filtered spectra is given in
\markcite{FGS97}Fardal et al. (1997).

\section{Ly$\alpha$ Forest Cloud Models}

As stated previously, we neglect the possibility that the absorbers consist of
more than one thermal and ionization phase to concentrate on the effects of the
shape of the photoionizing radiation spectrum. We model the Ly$\alpha$ forest
absorbers using the photoionization code CLOUDY version 90
(\markcite{F96}Ferland 1996). The absorber clouds are modelled as
plane-parallel slabs of constant density, illuminated on both sides by the
ionizing radiation field.  We adopt a fiducial column density N(H~I) $=10^{15}$
cm$^{-2}$ and assume a metallicity $Z = 10^{-2} Z_\odot$. We adopt the spectral
shapes for the incident radiation discussed in \S 2.1 and vary the strength of
the incident radiation through the ionization parameter, $U = n_\gamma / n_H$,
where $n_\gamma$ and $n_H$ are the number densities of ionizing photons and
hydrogen respectively. Each curve in Figs. 2, 4, and 6 represents a set of
photoionization models paired with the corresponding spectrum shown in Figs. 1,
3, and 5, respectively. 

Although we focus this paper on the effect of the spectral shape, we have also
explored the effect of varying the parameters assumed above. For the spectral
shapes considered here, there is little change in the ratios (C~II/C~IV is
reduced by $\sim 10\%$ and Si~IV/C~IV is raised $\sim 10\%$) if N(H~I) is
increased by a factor of 10, as long as the temperatures for the clouds are
assumed to be consistent with the thermal equilibrium solution of the CLOUDY
model. An increase (or decrease) by a factor of 3 in the assumed metallicity
has a slightly larger effect. The C~II/C~IV fraction is reduced $20-30\%$ with
a factor of 3 reduction in metallicity and raised $8-15\%$ by a factor of 3
increase in metallicity. The ratio Si~IV/C~IV is raised as much as $10\%$ and
reduced by $6\%$ or less with the same changes in metallicity. Constraints on
the H~I columns of the clouds preclude large reductions in the assumed density
for one-phase photoionization models, to avoid excessive length scales for the
clouds and to ensure that contributions to the baryon density from these N(H~I)
$\gsim 10^{15}$ cm$^{-2}$ absorbers do not exceed BBN limits. 

We normally allow the temperature to be that solved for by CLOUDY, assuming
thermal equilibrium between photoelectric heating and radiative cooling. If the
clouds are actually cooler, for example if they have expended some energy in
expansion, or if they retain thermal memory of earlier epochs when Compton
cooling off the microwave background was important, the ratios Si~IV/C~IV vs.
C~II/C~IV may be increased by as much as a factor of 2 at low values of
C~II/C~IV (high values of U). If, as suggested by \markcite{HRS97}Haehnelt,
Rauch, \& Steinmetz (1997), the temperatures of the clouds are larger than that
expected from photoionization thermal equilibrium, the Si~IV fraction is more
temperature-sensitive than the C~IV fraction, and Si~IV/C~IV is decreased
further relative to C~II/C~IV. These higher temperatures, suggested as a way to
increase the apparent thickness of the clouds, also imply much smaller
metallicities, as the C~IV fraction is decreased much less compared to the H~I
fraction by higher temperatures. For example, adopting the $J_{Q,f}$ spectrum
and assuming $U = 10^{-2}$, we find that increasing the cloud temperature from
25,000~K to 70,000~K decreases the C~IV fraction by less than 20\% but reduces
the H~I fraction (and inferred metallicity) by a factor of 8. 

There is a further effect if the assumed temperature of the cloud is increased
to the point that collisional ionization of H~I becomes important.  In this
case, He~II/H~I may be larger than that assuming pure photoionization only,
N(He~II) may exceed $1 / \sigma_{HeII}$, and the cloud may become
self-shielding.  Even if this is not the case for a cloud with N(H~I) $=
10^{15}$ cm$^{-2}$, this becomes increasingly likely for greater values of
N(H~I), so that radiative transfer within the individual clouds becomes
important. In Fig. 7 we show the results of models that explore some of these
temperature effects but defer a complete discussion to a later paper. 

\section{Results}

\subsection{The Metagalactic Radiation Background}

Figure 2 summarizes the results of our photoionization models which assume
different sources of photoionizing radiation for the metagalactic background.
The solid and dotted curves both assume photoionization by sources that possess
the same power-law spectral shape ($\alpha_s = 1.8$) but differ in the
treatment of the filtering of their radiation by intervening clouds (see Fig.
1). Models incorporating a pure-AGN spectrum, $J_{Q,f}$, indicate that
the Si~IV/C~IV ratio may be enhanced by almost a factor of 2 over the simpler
model for the absorbers with C~II/C~IV ${}^>_\sim 0.1$. Even in this regime, Si
is still overabundant relative to C, as previously argued by
\markcite{SC96}Songaila \& Cowie (1996a).  However, a factor of two enhancement
may be sufficient. If the absorbers are photoionized, an increasing C~II/C~IV
ratio is directly related to a decreasing ionization parameter (Figure 2). At
high C~II/C~IV (low U) there are few high-energy photons available to ionize
the higher ionization stages of silicon.  As a result, the enhanced ionization
of Si~III by He~II Ly$\alpha$ photons emitted by clouds is enough to increase
the ratio of Si~IV/C~IV over that of the broken power-law models. Compared to
more realistic calculations with cloud re-emission, models for the spectral
shape of the metagalactic background with a simple break at 4 Ryd underestimate
the number of high-energy photons in the metagalactic background. At low
C~II/C~IV (high U), the Si~IV fraction is strongly depleted by the increased
amount of high-energy photons in the realistic spectrum. 

A metagalactic ionizing background dominated by stellar sources, $J_{MB} =
J_{SQ,f}$, increases the Si~IV/C~IV ratios at C~II/C~IV $>$ 0.1 by $20-50\%$
over that of the $J_{MB} = J_{Q,f}$ case.  Once again, using a more realistic
spectrum increases the Si~IV/C~IV ratios by almost a factor of 2 compared to
spectra with simple attenuation of He~II ionizing photons. As in
the pure AGN case, the simple representation overestimates the corresponding
intensity at higher energies. To be compatible with observed Si~IV/C~IV ratios
if C~II/C~IV $>$ 0.1, the relative abundance of Si must be increased by
approximately a factor of 2.  In all cases, the large Si~IV/C~IV ratios in
absorbers with C~II/C~IV $<$ 0.1 are difficult to explain in the context of
single-phase models unless Si/C is enhanced by an order of magnitude. Chemical
evolution models with Si/C $\geq 3$ (Si/C)$_{\odot}$ are unrealistic, even if
the nucleosynthesis is dominated by massive stars (Woosley \& Weaver 1995).

\subsection{A Photoionizing Background Dominated by Local Sources}

One resolution to the puzzle of large Si~IV/C~IV at small C~II/C~IV may be that
the ionizing radiation incident on the absorbers is dominated by stellar
sources of radiation. As Fig. 4 shows, if the ionizing spectrum is entirely
stellar, very large ratios of Si~IV/C~IV are always possible due to the steep
dropoff in radiation more energetic than 45 eV. In practice, however, as Figs.
3 and 4 show, if the metagalactic background is primarily due to quasars
($J_{MB} = J_{Q,f}$), even if $J_{L}/J_{MB} = 20$, high Si~IV/C~IV ratios at
low C~II/C~IV require Si enhancements of a factor of 10. 

If the metagalactic background is dominated by stellar sources ($J_{MB} =
J_{SQ,f}$), large enhancements of Si may be unnecessary (see Figs. 5 and 6).
For many absorbers with C~II/C~IV $>$ 0.1, no overabundance of Si is necessary
if $J_{L}/J_{MB} = 3-10$.  At low C~II/C~IV, $J_{L}/J_{MB}$ must exceed $10$ if
enhancements in Si exceeding 2 are to be avoided.  As our order of magnitude
estimates in \S 2.1 indicate, this would require that the absorbers lie
within 30 kpc of starburst galaxies.  However, this situation is quite feasible,
given realistic constraints on absorber size and metal transport distances.

\subsection{Non-Equilibrium Temperatures and Non-Uniform Radiation Fields} 

As Figs. 7a and 7b show, if the temperatures of the clouds are cooler than the
photoionization thermal equilibrium temperatures
(\markcite{FG97}Ferrara \& Giallongo 1997;
\markcite{Z97} Zhang et al. 1997), 
much higher ratios of
Si~IV/C~IV are possible for low C~II/C~IV (high U) clouds.  These high U clouds
are precisely the clouds for which the photoionization thermal equilibrium
temperature is highest ($T \approx 40,000$~K).  The models denoted by the
dotted curves in Figs. 7a and 7b assume $T=15,000$~K, which may not be
compatible with the observed line widths, and which are cooler than is usually
assumed for Ly$\alpha$ forest clouds. If higher temperatures are associated
with the clouds, for example the dashed curves in Figs. 7a and 7b, which assume
$T = 50,000$~K, it is more difficult to account for the ratios in a one-phase
model. 

Another way to enhance Si~IV/C~IV at low C~II/C~IV, as suggested by
\markcite{SC96}Songaila \& Cowie (1996a) and \markcite{S97}Savaglio et
al. (1997), is to assume that no photons above 4 Ryd are present in the incident
ionizing spectrum. This is the limiting case of a cloud embedded in a region
where He II I-fronts have not yet overlapped. The Si~IV/C~IV ratios in Fig. 8
are not as high as those for the limiting case when all incident radiation is
starburst radiation, since radiation between 45 eV and 54 eV is included, which
will ionize Si~IV to Si~V. This effect is seen in comparing models which assume
pure AGN and starburst/AGN sources in Fig. 8. As Fig. 8 also shows, a careful
treatment of the propagation of these He II I-fronts is necessary, 
particularly because most AGN possess x-ray emission with spectra that flatten
to approximately $\nu^{-1}$ above 0.3 keV.  We have shown that including
high-energy radiation with E $>$ 25 Ryd is sufficient to reduce Si~IV/C~IV
ratios substantially at low C~II/C~IV. 

\section{Discussion}

From the results of our one-phase photoionization models we draw the following
conclusions: 

1.  {\it Ionizing Radiation Field}.-- The metagalactic radiation field 
is likely to include both power-law (AGN) and stellar (hot-star) components.  
The Ly$\alpha$ forest absorbers with metal lines may also experience
a local radiation field from starburst galaxies within 50--100 kpc.  

2.  {\it Si/C Overabundance}.-- For plausible mixtures of stellar and AGN
spectra in the metagalactic background, our photoionization models produce
enhanced Si~IV/C~IV ratios, consistent with high-$z$ absorbers with Si/C 
$\approx 2$(Si/C)$_{\odot}$ at low ionization (C~II/C~IV $>$ 0.1). For absorbers
with C~II/C~IV $<$ 0.05, it is difficult to account for the high values of
Si~IV/C~IV unless Si/C $>$ 10(Si/C)$_{\odot}$, an unrealistically large value
for massive-star nucleosynthesis.  These systems may include photoionization
from local stellar sources as well as hot, collisionally ionized gas. 

3.  {\it Local Ionizing Sources}.-- If the radiation field incident on the
absorbers is dominated by a nearby starburst galaxy, the Si~IV/C~IV ratios are
further enhanced. If C~II/C~IV $>$ 0.1, no Si/C overabundance is necessary to
explain Si~IV/C~IV ratios if absorbers are within about 40~kpc of a starburst
galaxy {\it and} the background is dominated by stellar sources. Even if the
metagalactic background is dominated by AGNs, close proximity to a starburst
galaxy may reduce the needed Si/C overabundance to a factor 1.5.  When
C~II/C~IV $^<_\sim 0.05$, it remains very difficult to account for the high
values of Si~IV/C~IV with photoionized models, although models with locally
dominated radiation may only require Si/C enhancements of a factor of 2-3. 

4.  {\it Non-Overlapping He~II I-Fronts}.-- Around quasars whose ionizing
continua have steep spectral indices ($\alpha_s > 1.84$), the He~II I-fronts
will lie within the H~I I-fronts.  If many absorbers lie in regions
where all photons above 4 Ryd are attenuated (\markcite{SC96}Songaila \& Cowie
1996a), we obtain good agreement with almost all measurements of Si~IV/C~IV 
if Si/C is enhanced by a factor of 2--3.  However, including even a small
contribution of higher energy photons in the background increases the needed
Si/C overabundance to an order of magnitude for absorbers with the lowest
C~II/C~IV ratios. 

5.  {\it Temperature Effects}.-- If the absorbers are cooler than expected for
thermal equilibrium between photoelectric heating and radiative cooling, the
Si~IV/C~IV ratio is increased for a given C~II/C~IV ratio. At low C~II/C~IV,
this enhances Si~IV/C~IV by a factor of 5 in our photoionization models.
Conversely, if the absorbers are hotter, Si~IV/C~IV is lower for a given
C~II/C~IV ratio. 

\vspace{0.5cm}

While we have described several processes that might increase the Si~IV/C~IV
ratio, an important additional constraint on these possibilities is the fact
that the high ratios are preferentially found at $z > 3.1$.  
Recent observations (\markcite{B97}Boksenberg 1997) challenge
this interpretation, by finding high Si~IV/C~IV ratios in
absorbers at $z = 2-3$.  If, however, Si~IV/C~IV rises above
$z > 3.1$,
this argues for a
time dependence to whatever process increases this ratio. It is this property,
as well as the increased He~II absorption toward Q0302-003 at $z=3.28$
(\markcite{J94}Jakobsen et al. 1994), that
\markcite{SC96}Songaila \& Cowie (1996a) use to support their suggestion that
the absorbers lie in regions where He~II I-fronts have not overlapped at $z >
3.1$.  
From a higher resolution HST/GHRS spectrum of Q0302-003,
\markcite{H97}Hogan, Anderson, \& Rugers (1997)
find evidence for residual transmitted flux below
the He~II edge.  Their 95\% confidence upper limit, 
$\tau_{HeII} \le 3$, makes it less necessary to propose
that Q0302-003 lies in a region where He~II I-fronts have
not overlapped, but does not preclude the
suggestion that such regions existed at $z>3.1$.
If this is the case, radiation outside of He~III regions would largely
limit the range in ionization stages for carbon to II-IV, and for silicon to
II-V.  In AGN, however, the large fluxes of unattenuated photons above 45 eV
can ionize Si~IV and C~III, so that
high observed Si~IV/C~IV ratios would be difficult to
explain for large U (small C~II/C~IV).  As a result, the spectrum below 4 Ryd
must be dominated by stellar sources. This may be difficult to achieve with the
known quasar luminosity functions at $z > 3.5$. In addition, soft x-ray
radiation from quasars, which is less attenuated by intervening clouds even in
the low He~III porosity case, may again make high Si~IV/C~IV, low C~II/C~IV
absorbers difficult to understand if they are solely photoionized by a
metagalactic background. 

New or improved measurements of He~II absorption may soon indicate whether
He~II I-fronts have overlapped well before $z = 3.1$. In that case, a sharp
change in the shape of the metagalactic radiation field may be less plausible.
Still, in general, the trend is likely to be  more attenuation of photons above
the 4 Ryd limit with increasing redshift. Other time dependent effects that
enhance the ratio of Si~IV/C~IV are possible.  Higher abundances of Si relative
to C are associated with metal yields from the most massive stars. However, in
a flat universe with $h=0.75$, the age of the universe exceeds $10^9$ years by
$z = 3.1$, time enough for lower mass stars to enrich the gas with carbon.  For
example, if multiple supernovae eject metal-enriched gas into the IGM, the
additional $4 \times 10^{8}$ years between $z=3.5$ and $z=2.5$ may make it more
likely that the gas surrounding the high mass stars has been enriched with
carbon from a previous episode of star formation. 

The temperature effects we discuss in \S 4 may also have a time dependence.  If
the IGM has been uniformly enriched by a much earlier episode of Population III
star formation, and if the Ly$\alpha$ forest clouds with metals have their
origin in growing overdensities in the IGM, then the absorbers observed at
higher $z$ may have formed at an earlier epoch. If Compton cooling off the
cosmic microwave background was a dominant coolant at this epoch
(\markcite{MR94}Miralda-Escud\'e \& Rees 1994), 
the cloud may have retained
memory of this lower temperature.  This effect is probably not large, since
there is not a significant evolution in the observed line widths. 

Alternatively, one may resolve the puzzle of the high Si~IV/C~IV, low C~II/C~IV
absorbers by relaxing the single-phase model which we used to explore the
effects of radiative transfer. As we have emphasized in a previous paper
(\markcite{GSS94}Giroux et al. 1994), 
the production of heavy elements in QSO absorption
systems is naturally accompanied by hot gas due to supernovae and hot-star
winds. 

\acknowledgments

We thank Mark Fardal for the calculated metagalactic background radiation
spectra. We also thank G. Ferland for discussions about updated versions of
CLOUDY, and L. Cowie and A. Songaila for providing a summary of measurements
and upper limits on metal column densities in advance of publication. This work
was supported by the Astrophysical Theory Program at the University of Colorado
(NASA grant NAGW-766). 

\clearpage

\begin{deluxetable}{ll}
\tablewidth{37pc}
\tablecaption{Significant Multi-stage Metals}
\tablehead{
\colhead{Species}        & \colhead{$E_{th}$(Ryd) }
}

\startdata
C II, III, IV    & 24.38, 47.89, 64.49 \nl
N I, II, III, IV & 14.53, 29.60, 47.45, 77.47 \nl
O I, II, III     & 13.62, 35.12, 54.94 \nl
Ne I, II, III    & 21.57, 40.96, 63.46 \nl
Si II, III, IV, V& 16.35, 33.49, 45.14, 166.77 \nl
Al II, III       & 18.83, 28.45 \nl
S II, III, IV, V & 23.33, 34.83, 47.3, 72.68 \nl
Fe II, III, IV   & 16.19, 30.65, 54.8 \nl
\enddata

\end{deluxetable}

\begin{deluxetable}{lll}
\tablewidth{37pc}
\tablecaption{Absorber Distribution}
\tablehead{
\colhead{Range in $N_{17}$}  & \colhead{$A_i$ } & \colhead{$\beta_i$}
}

\startdata
$10^{-5}-10^{-3}$  & 0.132                    & 1.41  \nl
$10^{-3}-0.02$     & $6.31\times 10^{-3}$     & 1.85  \nl
$0.02-10^{5}$      & $2.9\times 10^{-2}$     & 1.46  \nl
\enddata

\end{deluxetable}

\clearpage

\figcaption{ Mean intensity, in relative units, versus energy in Ryd. All
curves have been normalized to the same values at $\nu = \nu_H$ (1 Ryd). Solid
curve assumes IGM-filtered sources with AGN ($\nu^{-1.8}$) spectrum at $z=3.4$.
Dotted curve assumes a mean intensity with $\nu^{-1.8}$ above $\nu=\nu_H$ (1
Ryd) and $\nu=\nu_{HeII}$ (4 Ryd), with decrease by a factor of 2 above
$\nu_{HeII}$, designed to give same ratio, $J$(4~Ryd)/$J$(1Ryd), as the
properly filtered case. Short-dashed curve assumes an IGM-filtered spectrum:
2/3 starburst (from Sutherland \& Shull 1997) and 1/3 AGN . Long-dashed curve
assumes $\nu^{-1.8}$ above $\nu=\nu_H$ and $\nu=\nu_{HeII}$, and a decrease by
a factor of 12 above $\nu_{HeII}$, to give same ratio, $J$(4~Ryd)/$J$(1Ryd), as
the mixed-source spectrum above. } 

\figcaption{ Ratios Si~IV/C~IV versus C~II/C~IV in photoionized models with
N(H~I)= $10^{15}$ cm$^{-2}$. Curves match the spectra of Fig. 1. Triangles are
from measured column densities of Si~IV, C~IV, and C~II (Songaila \& Cowie
1996a,b). Squares are measurements and upper limits from Songaila \& Cowie
(1996a,b). Diamonds are measurements and upper limits from Savaglio et al.
(1996).  Same data are also used in Figs. 4, 6, 7, 8 with model calculations
of Si~IV/C~IV versus C~II/C~IV. } 

\figcaption{ Mean intensity, in relative units, versus energy in Ryd. All
curves have been normalized to the same values at $\nu = \nu_H$. All curves are
mixtures where $J_{MB}$ is the IGM-filtered AGN spectrum shown Figure 1, and
$J_{L}$ is the unfiltered (``local'') stellar spectrum (Sutherland \& Shull
1997). Curves assume $J_{L} / J_{MB} = 0$ (solid), $5$ (dotted), $10$
(short-dashed), and $20$ (long-dashed). Dot-long-dashed curve assumes only
local stellar radiation, $J_{L}$. } 
 
\figcaption{ Ratios of Si~IV/C~IV versus C~II/C~IV in photoionized models with
N(H~I)= $10^{15}$ cm$^{-2}$. Curves match the spectra of Fig. 3. } 

\figcaption{ Mean intensity, in relative units, 
versus energy in Ryd, where all curves have been
normalized to the same values at $\nu = \nu_H$. All curves are mixtures where
$J_{MB}$ is the filtered stellar/AGN spectrum shown in Figure 1, and $J_{L}$ is
the unfiltered stellar spectrum from Sutherland \& Shull (1997).  Curves assume
$J_{L} / J_{MB} = 0$ (solid), $3$ (dotted), $5$ (short-dashed), $10$
(long-dashed), and $20$ (dot-dashed). } 

\figcaption{ Ratios Si~IV/C~IV versus C~II/C~IV in photoionized models with
N(H~I)= $10^{15}$ cm$^{-2}$. Curves match the spectra of Fig. 5.  
Dot-long-dashed curve assumes only local stellar radiation, $J_{L}$. } 

\figcaption{ (a) Effects of varying cloud temperature.  All cloud models assume
a metagalactic background with quasar sources only, $J_{MB} = J_{Q,f}$. Solid
curve adopts the equilibrium temperature computed by CLOUDY for a given
ionization parameter and spectral shape (varies between $T = 16,500$ K and $T =
45,000$ K.) Dotted curve assumes constant $T = 15,000$ K for all models; dashed
curve assumes constant $T = 50,000$~K for all models. (b) Same as 7a, except
background includes both quasars and hot stars, $J_{MB} = J_{SQ,f}$. } 

\figcaption{ Ratios Si~IV/C~IV versus C~II/C~IV in an IGM in which He~II
ionization fronts  have not overlapped. Solid curve repeats models for the
filtered  starburst/AGN spectrum in Figure 2. Dotted assumes filtered
starburst/AGN spectrum with all photons with $E > 4$ Ryd removed. Long-dashed
curve assumes $\nu^{-1.8}$ AGN spectrum with all photons with $E > 4$ Ryd
removed. Short-dashed curve assumes filtered starburst/AGN spectrum with $4 < E
< 25$ Ryd photons removed. } 

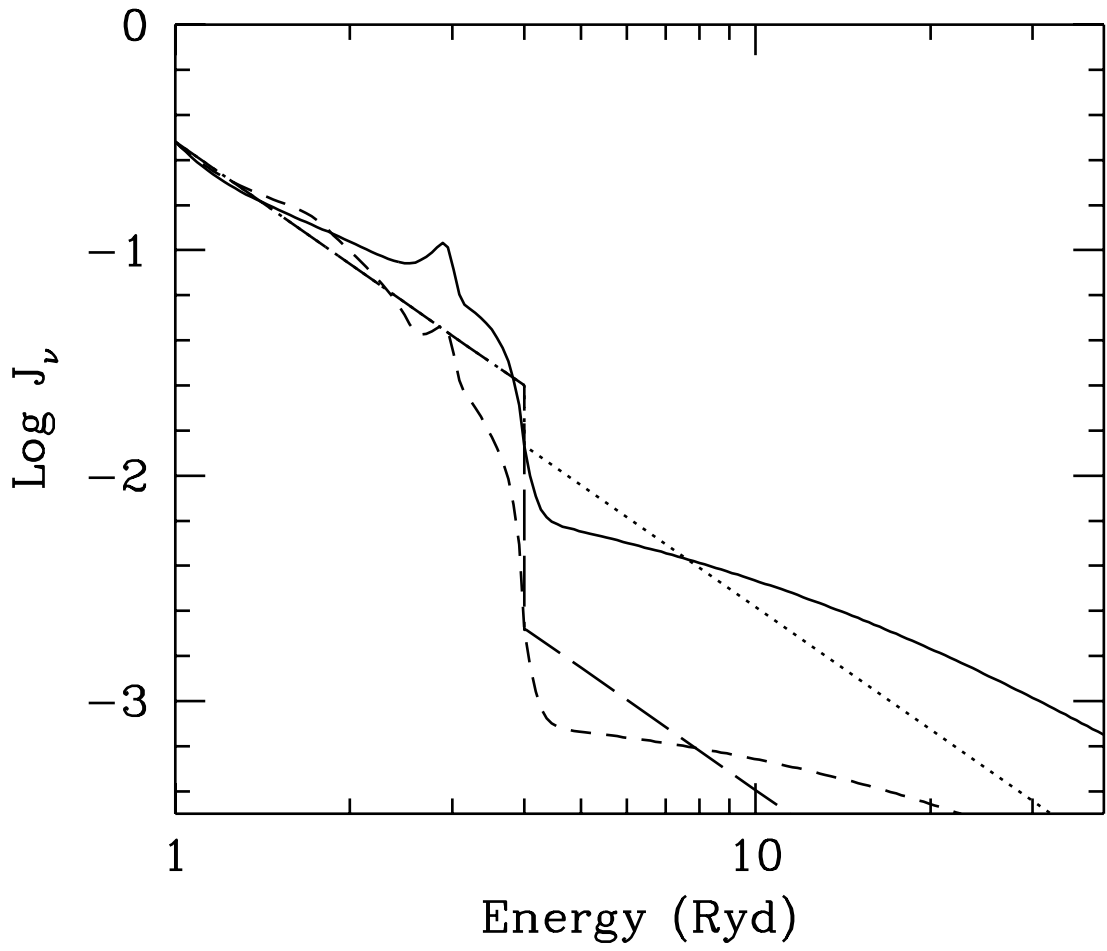
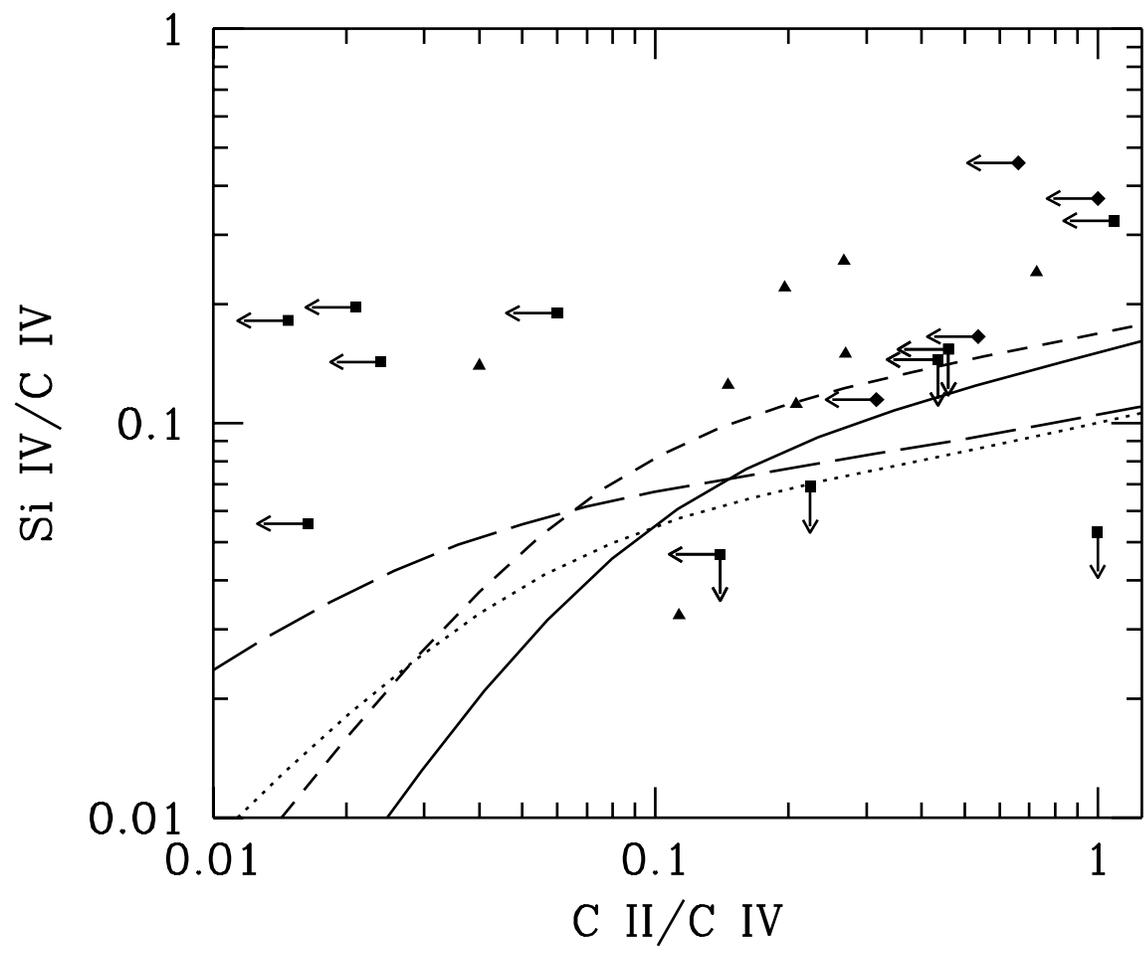
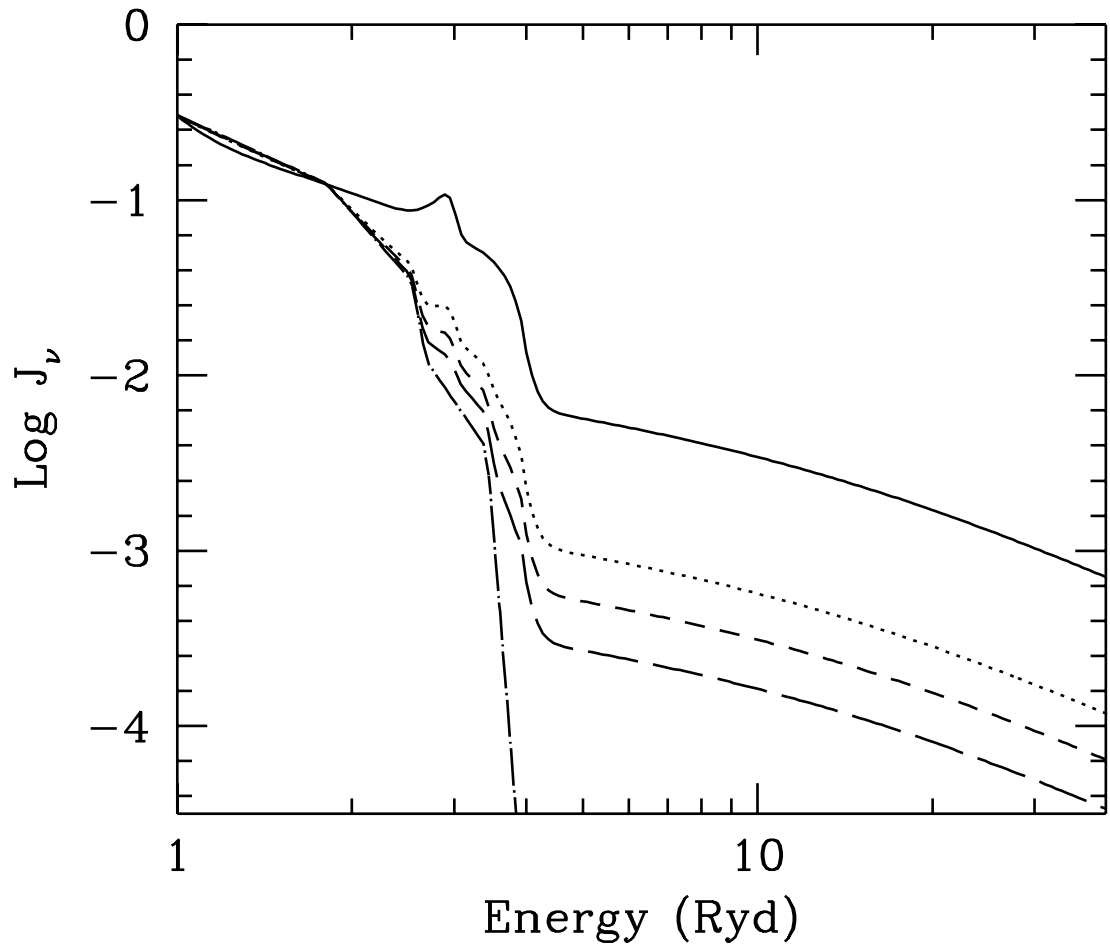
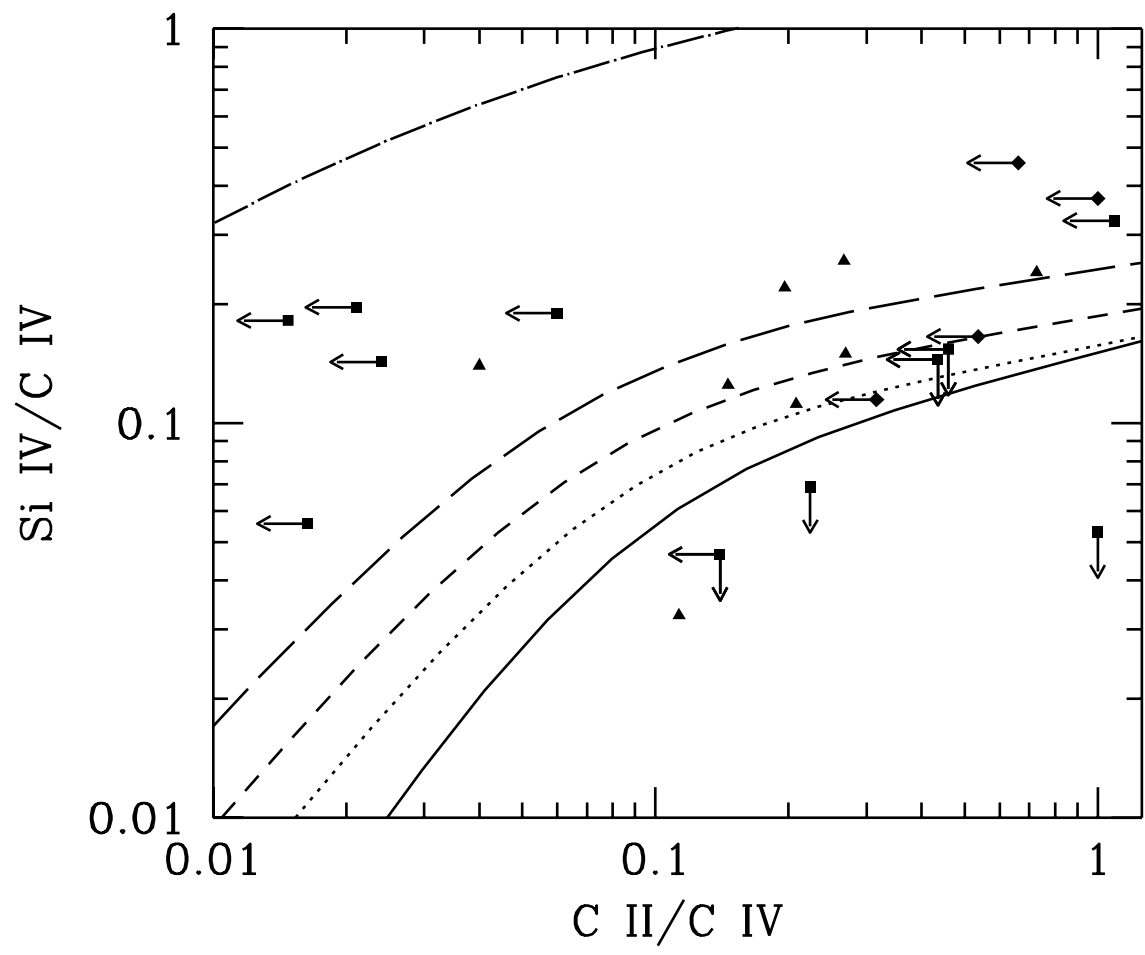
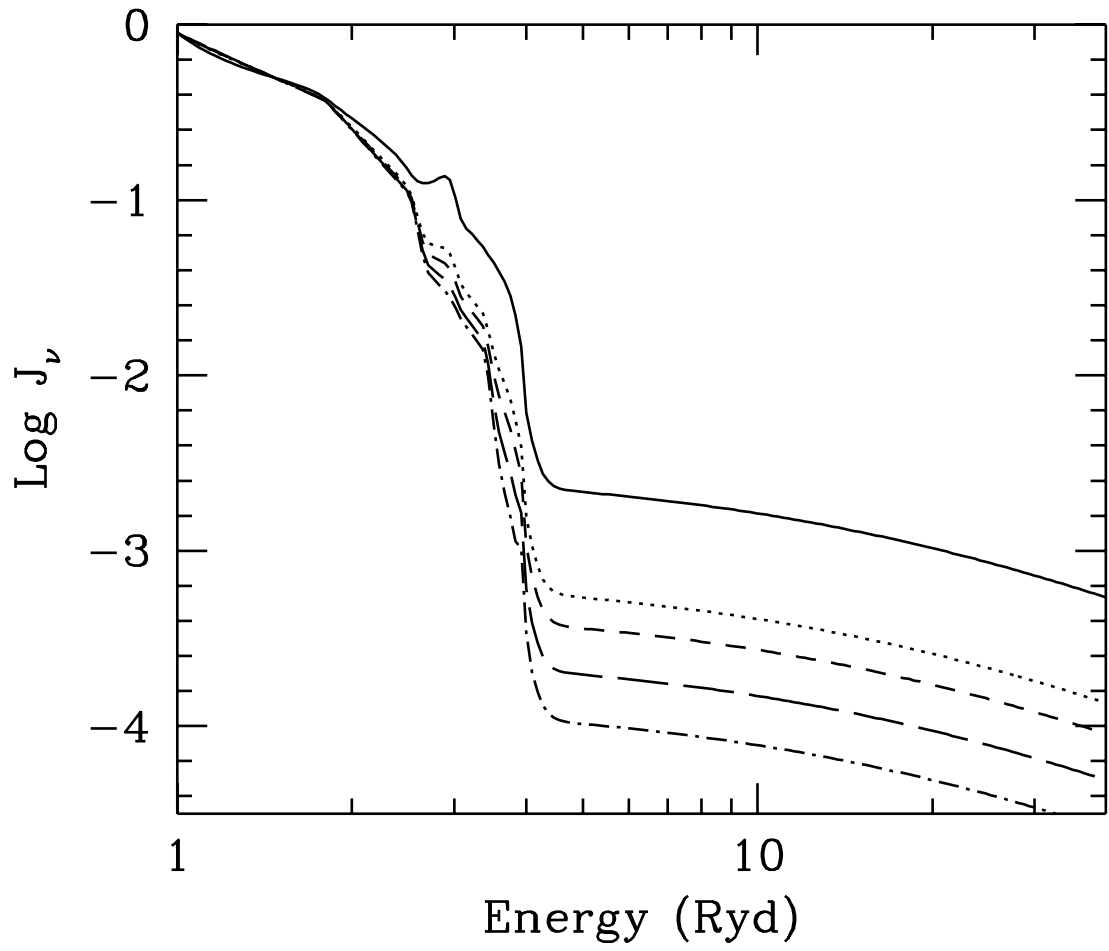
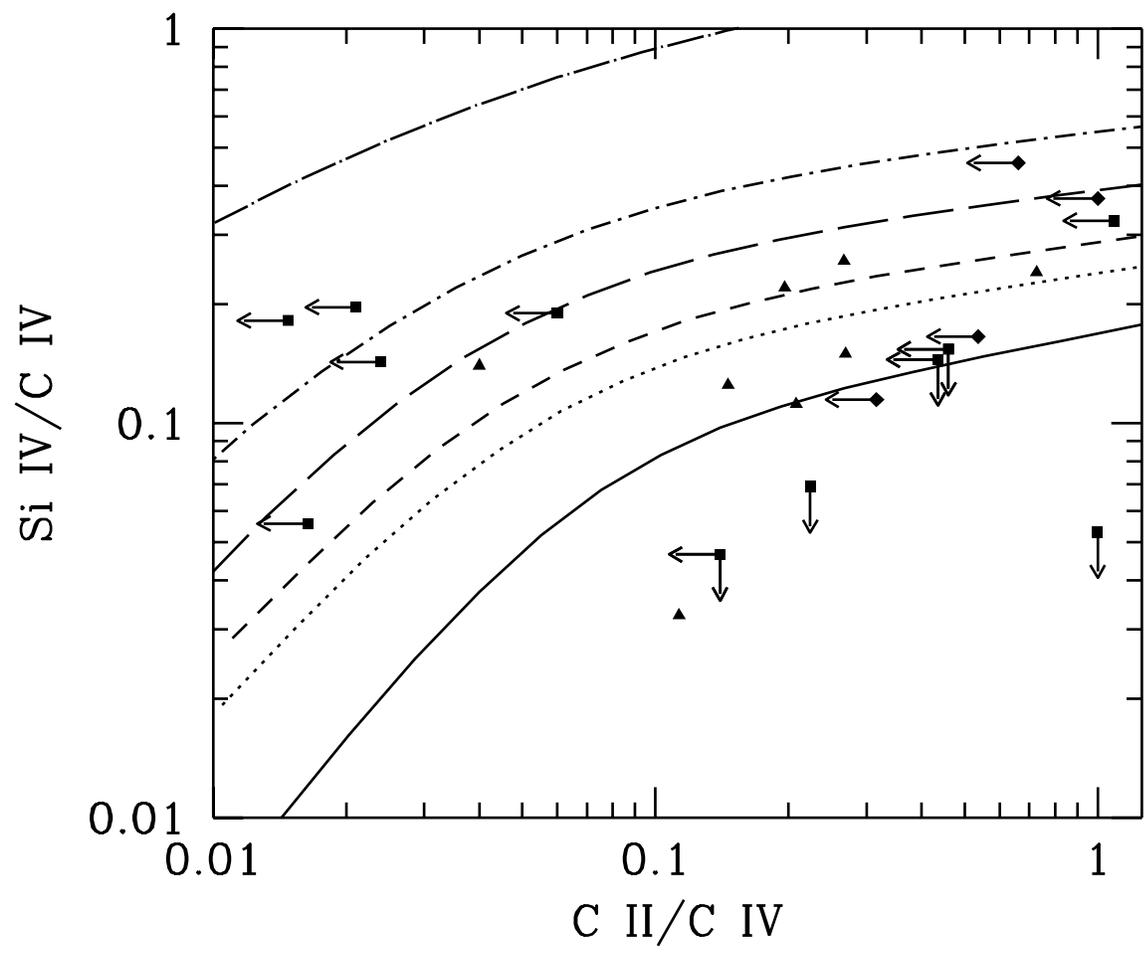
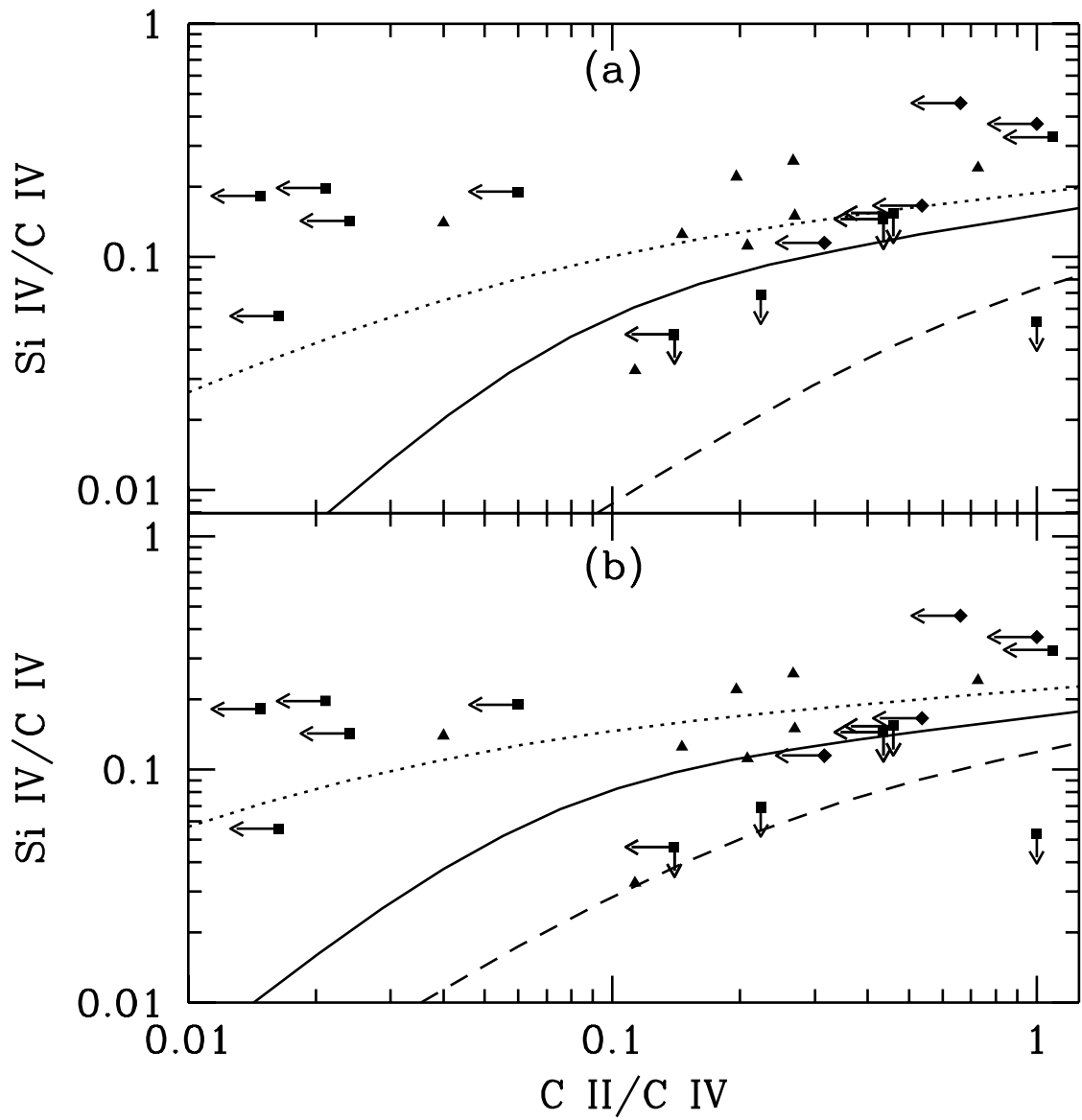
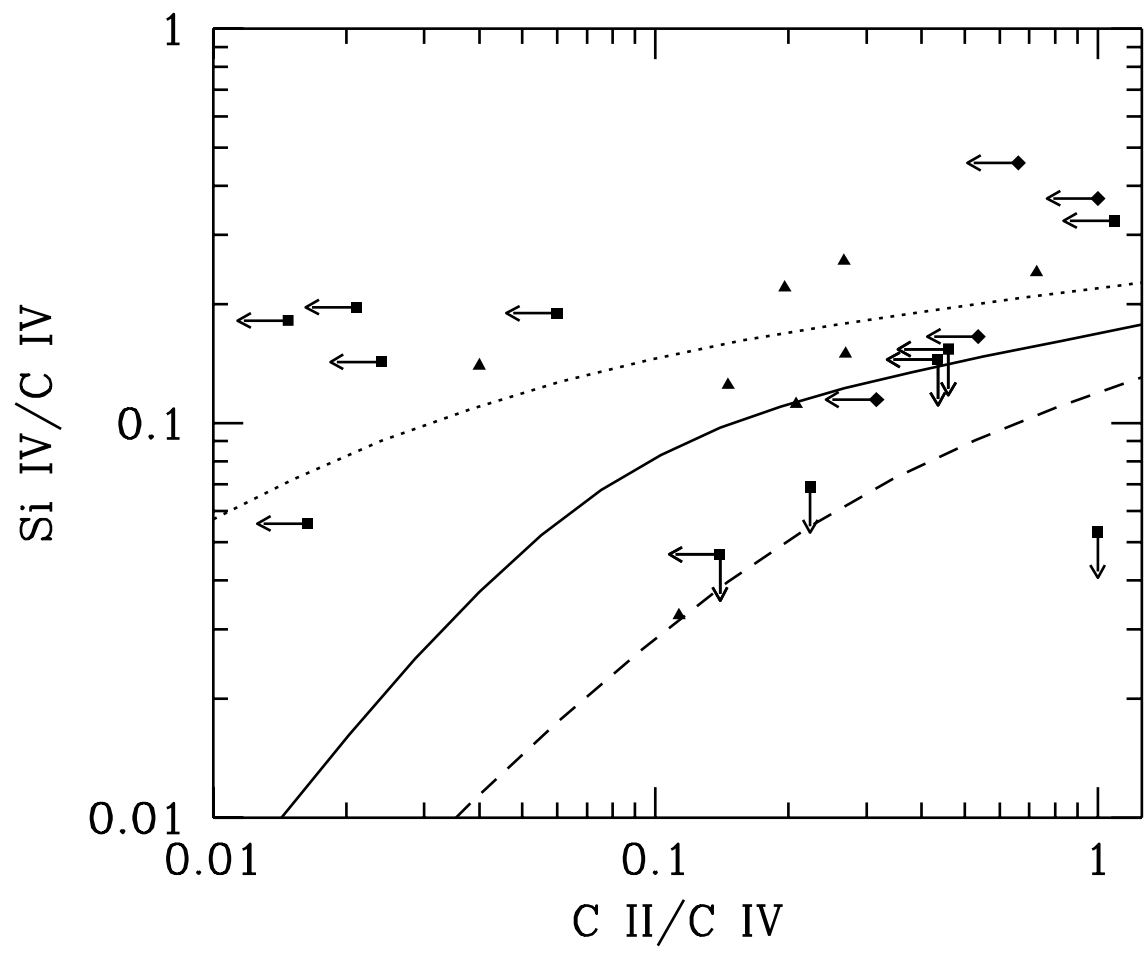
\end{document}